\documentclass{emulateapj}

\newcommand{\eh}[1]{\,\mathrm{#1}}

\newcommand{\dg}{^{\circ}}

\newcommand{\pcnt}{$\eh{\%}$}

\renewcommand{\epsilon}{\varepsilon}

\newcommand{\tss}[1]{\textsuperscript{#1}}

\shorttitle{Observations of the Blazar 3C 66A with the MAGIC Telescopes in
Stereoscopic Mode}
\shortauthors{Aleksi\'c et al.}

\begin{document}


\title{Observations of the Blazar 3C 66A with the MAGIC Telescopes in
Stereoscopic Mode}

%\input{authors_last_ApJ_2010_07_02.tex}

% authors 1.9.2010  Format ApJ (added Cossio 2.09., Thom 14.09)
%
\author{
J.~Aleksi\'c\altaffilmark{a},
L.~A.~Antonelli\altaffilmark{b},
P.~Antoranz\altaffilmark{c},
M.~Backes\altaffilmark{d},
J.~A.~Barrio\altaffilmark{e},
D.~Bastieri\altaffilmark{f},
J.~Becerra Gonz\'alez\altaffilmark{g,}\altaffilmark{h},
W.~Bednarek\altaffilmark{i},
A.~Berdyugin\altaffilmark{j},
K.~Berger\altaffilmark{g}
% ...
}
\altaffiltext{a} {IFAE, Edifici Cn., Campus UAB, E-08193 Bellaterra, Spain}
\altaffiltext{b} {INAF National Institute for Astrophysics, I-00136 Rome, Italy}
\altaffiltext{c} {Universit\`a  di Siena, and INFN Pisa, I-53100 Siena, Italy}
\altaffiltext{d} {Technische Universit\"at Dortmund, D-44221 Dortmund, Germany}
\altaffiltext{e} {Universidad Complutense, E-28040 Madrid, Spain}
\altaffiltext{f} {Universit\`a di Padova and INFN, I-35131 Padova, Italy}
\altaffiltext{g} {Institute de Astrof\'{\i}sica de Canarias, E-38200 La Laguna, Tenerife, Spain}
\altaffiltext{h} {Depto. de Astrof\'{\i}sica, Universidad de La Laguna, E-38206 La Laguna, Spain}
\altaffiltext{i} {University of \L\'od\'z, PL-90236 Lodz, Poland}
\altaffiltext{j} {Tuorla Observatory, University of Turku, FI-21500 Piikki\"o, Finland}
% ...


\email{ksaito@mppmu.mpg.de, klepser@ifae.es}



\begin{abstract}
We report new observations of the intermediate-frequency peaked BL Lacertae
object 3C~66A with the MAGIC telescopes. 
The data sample we use were taken in 2009 December and 2010 January,
and comprises 2.3 hr 
of good quality data in stereoscopic mode. In this period, we find a significant
signal from the direction of the blazar 3C~66A. 
The new MAGIC stereoscopic system is shown to play an essential role for the separation between 
3C~66A and the nearby radio galaxy 3C~66B, which is at a distance of only $6^\prime$.
The derived integral flux above $100\eh{GeV}$ is 8.3\pcnt\ of Crab Nebula flux and 
the energy spectrum is reproduced by a power law of photon index 
$3.64 \pm 0.39_{\rm stat} \pm 0.25_{\rm sys}$. Within errors, this is
compatible with the one derived by VERITAS in 2009. From the spectra corrected
for absorption by the extragalactic background light, we only find small
differences between the four models that we applied, and constrain
the redshift of the blazar to $z < 0.68$.
\end{abstract}




\section{Introduction}

Blazars make up the majority of extragalactic sources of very high energy (VHE; 
$E > 100\eh{GeV}$) gamma rays. They are a subset of active galactic nuclei (AGNs), 
and consist of BL Lacertae (BL Lac) objects and flat-spectrum radio-loud quasars.
The general framework to explain the gamma ray emission is that they are
produced by charged particles which are accelerated in a relativistic jet.
These jets are powered by gas accretion into a central supermassive black hole
and are perpendicular to the accretion disc.
When the jet is directed to us, the energy and flux of gamma-rays are boosted by the relativistic 
beaming effect 
\citep[e.g.,][]{bla78, urr95}. 

Generally, the spectral energy distribution of AGNs can be described by two broad bumps.
The lower energetic bump, at frequencies from radio to X-rays, is attributed to synchrotron
emission from nonthermal relativistic electrons in the jet. The other bump,
covering the X-ray to gamma-ray bands, could either be due to inverse Compton scattering 
of seed photons by the electrons \citep[leptonic model, e.g.,][]{mar92, der93, blo96, kra04} or due to 
hadronic interactions \citep[see, e.g.,][]{man93, mue01, mue03}. 

3C~66A was classified
as a BL Lac object by \citet{mac87}, based on its significant optical and
X-ray variability. The synchrotron peak of this source is located between 10$^{15}$ and
10$^{16}$ Hz \citep{per03}, therefore 3C~66A can also be classified as an
intermediate-frequency peaked BL Lac object (IBL). 
The redshift of 3C~66A was determined to be
$z = 0.444$ by independent authors \citep{mil78,lan93}.
However, their measurements are based on the detection of one single line.
Another observation of 3C~66A at a different spectral range was reported by \citet{fin08}, but no spectral 
feature was found, and a lower limit of the redshift was
derived to be 0.096. For the marginally
resolved host galaxy \citep{wur96}, a redshift of 0.321 was found.
Recently, through the investigation of the Large Area Telescope (LAT), on board the {\it Fermi Gamma-ray 
Space Telescope} ({\it Fermi}) satellite and VHE gamma-ray observations, upper limits for the 
redshift of 3C~66A were derived; 
$z=0.44$ \citep[][$2\,\sigma$ confidence level]{pra10} and $z=0.58$ \citep{yan10b}. 

Several gamma-ray observations of 3C~66A were performed since the 1990s. With
the {\it EGRET}
satellite, a GeV gamma-ray emission (3EG J0222+4253) was associated with 
3C~66A \citep{har99}. However, due to the large {\it EGRET} point-spread function (PSF), an influence by the 
nearby pulsar PSR J0218+4232 could not be excluded \citep{kui00}. The Crimean 
Astrophysical Observatory claimed detections of 3C~66A above $900\eh{GeV}$ with an integral 
flux of $(3\pm1)\times 10^{-11}\eh{cm^{-2} s^{-1}}$ \citep{ste02}. Later 
observations by HEGRA and Whipple reported upper limits of 
$F (>630\eh{GeV}) < 1.42 \times 10^{-11}\eh{cm^{-2} s^{-1}}$ \citep{aha00} and 
$F (>350\eh{GeV}) < 0.59 \times 10^{-11}\eh{cm^{-2} s^{-1}}$ \citep{hor04}, 
respectively. Additionally, the STACEE observation found a hint of signals at a 2.2 
significance level and derived upper limits of
$ < 1.0 \times 10^{-11}\eh{cm^{-2} s^{-1}}$ and $ < 1.8 \times 10^{-11}\eh{cm^{-2} s^{-1}}$ 
for thresholds of $147\eh{GeV}$ and $200\eh{GeV}$, respectively \citep{bra05}. 

Recent VERITAS observations of 3C~66A taken from 2007 September to 2008 January and 
from 2008 September to 2008 November, for a total of 32.8 hr, resulted 
in a detection in VHE gamma rays \citep{acc09}.
The energy spectrum was derived with a photon index of 
$\Gamma = 4.1\pm0.4_{\rm stat}\pm0.6_{\rm sys}$. 
The integral flux of the VERITAS observations above $200\eh{GeV}$ is 
($1.3\pm 0.1)\times 10^{-11}\eh{cm^{-2}\,s^{-1}}$ (6\pcnt\ of the Crab Nebula flux).  

3C~66A has been monitored by {\it Fermi}/LAT since 2008 August, covering the latter
part of the VERITAS observation. According to
\citet{abd09}, who reported 
the first 5.5 months of {\it Fermi}/LAT observations of 
3C~66A, the blazar showed a significant flux variability (a factor of 5--6
between the highest and lowest fluxes). The derived energy spectrum with the
photon index of $\Gamma = 1.98$ above $1\eh{GeV}$, in combination with the VERITAS
spectrum, indicates that the spectrum must soften above $100\eh{GeV}$.

MAGIC observed the sky region around 3C~66A from 2007 August to December, obtaining a total exposure
time after data quality cuts of 45.3 hr \citep{ali09b}.
These data revealed a significant VHE gamma-ray signal centered at
2\tss{h}23\tss{m}12\tss{s}, 43$^\circ$0$^\prime$7$''$.
This excess (named MAGIC~J0223+430) coincides within uncertainties with the position of a nearby, 
Fanaroff-Riley-I (FRI) type galaxy 3C~66B \citep[$z=0.0215$;][]{stu75}. Still,
judging from the skyplot alone, the probability of the emission to originate
from 3C~66A is 14.6\pcnt.
The energy spectrum of MAGIC~J0223+430
was reproduced by a single power law with the index of $\Gamma = 3.1\pm0.3$.
The integral flux above $150\eh{GeV}$ corresponded to 
($7.3\pm 1.5)\times 10^{-12}\eh{cm^{-2}\,s^{-1}}$ (2.2\pcnt\ of the Crab Nebula flux).
According to \citet{tav08}, the radio galaxy is also a plausible source of VHE 
gamma-ray radiation. Also, the recent MAGIC detection of IC
310 \citep{mar10}, a radio galaxy at a very similar redshift ($z=0.0189$)
indicates that 3C~66B might be feasible to explain all or part of the MAGIC detection from 2007.

\section{Observations} \label{sec_obs}

From mid 2009 August, 3C~66A went into an optical high state which was 
reported by the Tuorla blazar monitoring
program\footnote{http://users.utu.fi/kani/1m/index.html}. This outburst triggered new MAGIC 
observations. The optical flux in the $R$ band reached a maximum level of 
$\sim$$12\eh{mJy}$ in 2010 January, while the baseline flux in the historical 
data of the source is $\sim$$6\eh{mJy}$. 

The observations were carried out with the MAGIC telescopes located on
the Canary Island of La Palma (28.\hspace{-1.25mm}\degr8~N, 17.\hspace{-1.25mm}\degr8~W, $2220\eh{m\,a.s.l.}$). The
two $17\eh{m}$ diameter telescopes use the atmospheric Cherenkov imaging technique and
allow for measurements at a threshold as low as $50\eh{GeV}$ in normal trigger
mode. 

We observed the blazar 3C~66A in several time slots between 2009 September and
2010 January. However, the sky imaging CCD cameras that are used to
cross-check the telescope pointing ("starguider cameras") only became fully applicable to stereo
observations in early December. To allow for a high-confidence directional statement on the
arcminute scale, we therefore only used data taken after these upgrades, which
were 5.6 hr in total. Furthermore, we had to discard data with low event
rates, affected by the exceptionally bad weather conditions in that winter.
Finally, we had 2.3 hr of good quality data left after all quality cuts.
They were taken on six days
between 2009 December 5 and 2010 January 18, partly under low-intensity moon
light conditions.

The data were taken using the
false source tracking (wobble) method \citep{fom94}, in which the pointing
direction alternates every 20 minutes between two positions, offset by $\pm0.\hspace{-1.25mm}\dg4$ in RA from
the source. These wobble positions were chosen with respect to 3C~66A, but the small distance to
3C~66B (0.\hspace{-1.25mm}\degr01) allows equal judgment for both sources. The data were taken at zenith angles
between 13\degr\ and 35\degr.

\section{Data Analysis} \label{sec_ana}

For the analysis, only stereoscopic events triggered by both MAGIC
telescopes were used. They were analyzed in the MARS analysis framework
\citep{mor09}, taking advantage both of the advanced single-telescope
algorithms \citep[e.g.,][]{ali09a} and newly developed stereoscopic analysis routines.
These routines are at present still subject to some minor improvements and will
be discussed in more detail in a separate paper still in preparation, but are
shortly outlined in the following.

Combining monoscopic and stereoscopic strategies, the direction of gamma rays is calculated for each telescope
separately, using the random forest technique \citep{alb08b}, and later
combined with the projected crossing point of the image axes, with a weight depending on
the angle between the two shower images. Requiring a certain level of
agreement between the different estimates furthermore improves the resolution,
and also helps to reject the (less focused) hadron showers.
Similarly, an energy estimator is determined from look-up tables for each
telescope separately, and later combined to a common estimated energy.

The skymap generation, which is particularly important for the analysis of data
from the 3C~66A/B region, follows a two-step algorithm. The first step is to
generate an exposure model for the field of view in camera coordinates, for
the quality cuts that were applied in the
analysis. This is done by joining the distributions of photon-like
events from the two wobble positions, taking advantage of the fact that the
source, in relative camera coordinates, is on opposite sides for both wobble
sets.

The second step is the calculation of an expected background event
distribution in celestial coordinates, and
its comparison to the actual event distribution. Before that comparison, a
smearing with a Gaussian kernel is applied.
The significances are calculated following Equation (17) of
\citet{lim83}, taking into account the higher precision of the background estimation implied by
the above modeling. 

The performance of the analysis software was optimized and checked with
contemporaneous Crab Nebula data and MC. The Crab Nebula spectrum could be
analyzed down to about $50\eh{GeV}$, fully covering the range of the spectrum
presented in the next paragraph. The
achieved angular resolution, defined as the $\sigma$ of a two-dimensional
Gaussian function, is around 0.\hspace{-1.25mm}\degr1\ at $100\eh{GeV}$ and
approaching 0.\hspace{-1.25mm}\degr065\ at higher energies. This $\sigma$ defines the radius in
which 39\pcnt\ of all photons of a point source are contained. The systematic uncertainty on the
direction reconstruction is a product of the telescope pointing uncertainty
and possible biases that occur in the reconstruction algorithms. The latter
can be caused by irregularities in the shower images, such as missing camera
pixels, inhomogeneous noise from stars in the field of view, or
imperfections in the data acquisition electronics.
Both the total
pointing deviation and the telescope pointing precision of MAGIC were always
monitored over the years \citep{bre09, ale10}, and along with studies of
contemporary stereo data of known direction lead to an estimate of the maximal systematic
stereoscopic pointing uncertainty of 0.\hspace{-1.25mm}\degr025.

We also used the publicly accessible {\it Fermi}/LAT
data\footnote{http://fermi.gsfc.nasa.gov/}
to investigate the status of the source in the GeV 
energy range during the MAGIC observation period. The {\it Fermi} data were analyzed using the public 
software package LAT Science Tools v9.15.2, including the Instrument Response File 
P6\_V3\_DIFFUSE, and galactic, extragalactic and instrumental background models.

\section{Results} \label{sec_res}

\begin{figure}
\plotone{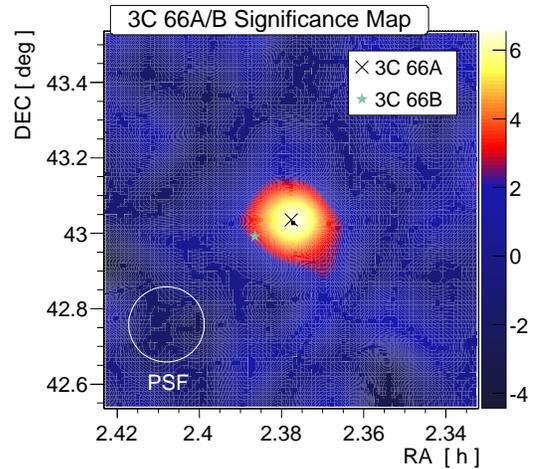}
\caption{MAGIC significance skymap of the region around 3C~66A/B for events with energies
above $100\eh{GeV}$.
\label{fig1}}
\end{figure}

\begin{figure}
%\epsscale{.80}
\plotone{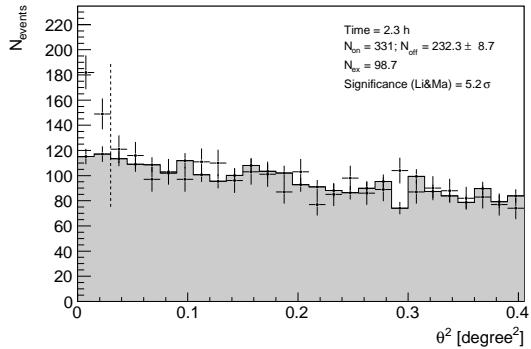}
\caption{Distribution of squared angular distances between photon directions
and the position of 3C~66A ($\theta^2$) for events with energies
above $100\eh{GeV}$. The OFF data are taken from three
positions that are symmetrical with respect to the telescope pointing
directions. 
\label{fig2}}
\end{figure}

Figure~\ref{fig1} shows a skymap of the observed region above $100\eh{GeV}$.
The significance of the excess at the location of 3C~66A is $6.4\eh{\sigma}$.
We cross-checked the detection also by investigating the
distribution of squared angular distances ($\theta^2$) between photon
directions and the assumed source position.
The expected background is extracted from corresponding $\theta^2$ plots done with respect to other sky positions
at similar
distance from the pointing direction. Comparing the data with this expectation 
we find a significance of $5.2\eh{\sigma}$ (see Figure~\ref{fig2}). The difference in
significance can be attributed to the different
integration procedure of
signal and background in the skymap, which generally leads to a slightly better
background estimation and therefore a higher significance.

We also analyzed the data taken with and without moon light separately to find 
possible effects from the higher thresholds of individual camera pixels.
However, we could not find a clear tendency beyond the statistical errors and
thus decided to use all the data for the analysis.

Unlike in the 2007 observations of this sky region, the emission peak this time is
clearly on top of 3C~66A. The fitted center of gravity
of the excess (small black square in Figure~\ref{fig1}) is at a distance of
$0.\hspace{-1.25mm}\dg010 \pm 0.\hspace{-1.25mm}\dg023\, \mathrm{(stat.)} \pm 0.\hspace{-1.25mm}\dg025\, \mathrm{(sys.)}$
from 3C~66A, and $0.\hspace{-1.25mm}\dg108 \pm 0.\hspace{-1.25mm}\dg023\,\mathrm{(stat.)} \pm
0.\hspace{-1.25mm}\dg025\, \mathrm{(sys.)}$ from 3C~66B. While being compatible with the former,
the statistical rejection power for the emission to emerge from the radio
galaxy 3C~66B corresponds to 4.6 standard deviations. Even considering the unlikely case
of a systematic offset exactly toward the blazar, the
rejection significance of 3C~66B is at least $3.6\eh{\sigma}$. 
These numbers were confirmed
by a second analysis with independent data quality selection and cut
optimization procedures. The same result is found even when
the photon direction is taken only from the projected crossing point of the two shower
axes. We therefore conclude that the signal we see this time emerges from the blazar
3C~66A.

It shall be mentioned that this result is a clear
merit of the angular resolution and background rejection of the new
stereoscopic system. In fact, if we compare the above stereo directional
reconstruction algorithm to the MAGIC-I algorithm alone, we find basically the
same result, but the
statistical error of the fitted source position increases roughly by a factor of two.
Consequently, the rejection significance of 3C~66B would be less than 2
standard
deviations, and the total detection significance would be below 5 standard 
deviations.

\begin{figure}
%\epsscale{.80}
\plotone{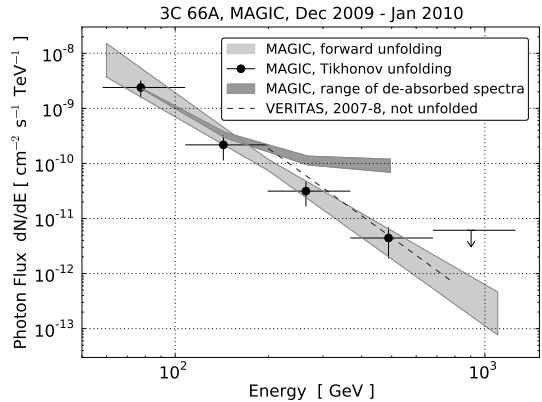}
\caption{Observed and EBL-corrected (de-absorbed) differential energy spectra of 3C~66A in the period of 2009 December and 2010 January. The
light-shaded area indicates the $1\eh{\sigma}$ range of the observed power law spectrum
gained by forward unfolding (see the text), the
crosses are from the unfolding after \citet{tik79} for comparison. The
dark-shaded area is the spread of the de-absorbed, mean flux values obtained by the
four applied EBL models, assuming the redshift of $z$=0.444. The VERITAS
(observed) spectrum after \citet{acc09} is shown for comparison.
\label{fig3}}
\end{figure}

The energy spectrum of 3C~66A was derived using four different unfolding
algorithms \citep{alb07} which correct for efficiency, smearing and biasing
effects in the energy response of the detector. The most conservative of these
methods is the so-called forward
unfolding, in which essentially a spectral shape is assumed a priori, and its
parameters are adjusted by iteratively folding the assumed spectrum with the response function
until the predicted distribution of estimated energies matches optimally the
actually measured distribution. With all unfolding methods, we found that the data are well compatible with a
power law of the form

\begin{equation}
\frac{dF}{dE} = K_{200}\left(\frac{E}{200\,\mathrm{GeV}}\right)^{-\Gamma},
\end{equation}

with a photon index $\Gamma = 3.64 \pm 0.39_{\rm stat} \pm 0.25_{\rm sys}$ and
a flux constant at $200\eh{GeV}$ of $K_{200} = 9.6 \pm 2.5_{\rm stat} \pm
3.4_{\rm sys}\, \times 10^{-11}\eh{cm^{-2}\,s^{-1}\,TeV^{-1}}$. 
The integral flux above $100\eh{GeV}$ corresponds to $(4.5\pm1.1)\times
10^{-11}\eh{cm^{-2}\,s^{-1}}$ 
(8.3\pcnt\ Crab Nebula flux). Here, the
parameters and statistical errors are taken from the forward unfolding,
while the systematic errors reflect the variations among the other
unfolding algorithms, plus several standard uncertainties discussed
in \citet{alb08a}. The systematic flux uncertainties add up to 36\pcnt\ in
total. Figure~\ref{fig3} displays the function we fitted through forward
unfolding, and spectral points derived using the Tikhonov unfolding method
\citep{tik79}. 

Due to the shortness of our observation, we cannot discuss flux variability
with these data. However, comparing the flux to the one from our previous
observation of the 3C 66A/B region confirms the VERITAS report of 3C 66A
being a variable source in general.

We also analyzed the {\it Fermi} data from the same time period. The flux
variability we found in a week-to-week light curve is not significant.
Given the statistical uncertainties of the light curve, we would be
sensitive on $3\eh{\sigma}$ level to flux variations of 60\pcnt\ or greater, and
conclude the variability in the days we observed must be less than
that. The averaged flux above $200\eh{MeV}$ is roughly comparable to the
averaged flux over the first 5.5 months \citep{abd09}, and lower than that seen
in 2008 October, when a strong TeV flare was observed by VERITAS. A
single power-law model can reproduce the source spectrum, and the photon
index is compatible with the one found in \citet{abd09}, indicating no
significant change in the overall spectral shape.

\section{Discussions and Conclusions} \label{sec_dis}

MAGIC observed the 3C~66A/B region in 2009 December and 2010 January, during
an 
optical active state of 3C~66A and detected a clear VHE gamma-ray signal.
The excess coincides with the position of 3C 66A, and we rule out the emission to come
from 3C~66B at a confidence level of $3.6\eh{\sigma}$. This detection does not
contradict the earlier MAGIC detection, though, which favored 3C~66B as the VHE
source. On the one hand, because the observation time of 2.3 hr would be
too short to detect the VHE emission of 3C~66B, if on a similar flux level
as in 2007, and on the other hand, because its flux may be even lower than
before. In fact, 3C~66A might have to
be in a low flux state in order not to outshine the comparably weak emission
from 3C~66B at this close distance of about $1\eh{\sigma}$ of the PSF of the MAGIC
telescopes.

The obtained energy spectrum is softer than in the previous MAGIC detection
($\Gamma = 3.10 \pm 0.31_{\rm stat} \pm 0.2_{\rm sys}$) and
compatible with the VERITAS spectrum of 3C~66A.
Compared to VERITAS, the MAGIC measurement has a lower threshold and the spectrum is 
extending to well below $100\eh{GeV}$. The flux level of 8.3\pcnt\ Crab Nebula
flux is similar to the one reported by VERITAS (6\pcnt), and significantly
higher than in the previous MAGIC observation (2.2\pcnt).

The VHE photons produced at the source can be absorbed in the intergalactic
space by pair production with the low energy 
(UV to infrared) photons of extragalactic background light 
\citep[EBL;][]{ste92,hau01}. The amount of absorption depends on the energy and
redshift, and can be corrected for in the data, assuming a certain modeling of the
EBL density. Such a de-absorbed spectrum can be regarded as the spectrum we
would measure if there were no EBL. To derive a de-absorbed spectrum,
we tested several state of the art EBL models, namely, \citet{fra08}, the fiducial model in \citet{gil09}, \citet{kne10}, and \citet{dom10}. 
The EBL corrections were applied in the spectrum unfolding procedure (see
above), using the full covariance matrix to correctly calculate the errors. 
The spread of the differential, de-absorbed flux spectra, obtained with the
four models and assuming the redshift of $z=0.444$, is shown as the dark
shaded area in Figure~\ref{fig3}. 
The de-absorbed photon indices for the four EBL modelings are listed in Table~\ref{tbl1}. 
The differences between the de-absorbed spectra are very small, 
although the one corrected after \citet{kne10} is slightly harder than the
others. This also reflects the fact that also the predicted EBL shapes and densities are very
similar in the first three models, but the overall density in \citet{kne10} is somewhat
higher.

\begin{deluxetable}{lrr}
\tabletypesize{\scriptsize}
%\rotate
\tablecaption{EBL Corrected Indices\label{tbl1}}
\tablewidth{0pt}
\tablehead{
\colhead{Model} & \colhead{$\Gamma_{\rm int}$}
}
\startdata
\citet{fra08} & $2.57\pm0.68$  \\
\citet{gil09} & $2.61\pm0.67$  \\
\citet{dom10} & $2.59\pm0.68$  \\
\citet{kne10} & $2.37\pm0.70$  \\
\enddata
%\tablenotetext{*}{The fiducial model.
\end{deluxetable}

From most VHE emission models, the de-absorbed spectrum is expected not to be
concave, i.e., rising toward higher energies. This can be tested both by comparing the
points of our own spectrum, but also by a comparison with the {\it Fermi} photon
index (1.98). The fact that we find our spectrum neither significantly
concave nor harder than in {\it Fermi} suggests that the assumed redshift of
$z=0.444$ does not contradict our observations.
In fact, we investigated the plausibility of the redshift,
assuming that the intrinsic spectrum is not expected to be exponentially
rising, and thus have a pileup, at highest
energies. This common method was previously used and described, for example, in \citet{maz07a, maz07b}.
Using the \citet{fra08} model and the likelihood ratio test between the "power
law" and "power law + pile-up" hypotheses, as described in the reference, we derive an upper limit on the redshift of 
$z<0.68$. 

The results derived in this paper demonstrate the 
advantages of the MAGIC stereoscopic system. Further MAGIC and other
gamma-ray observations of this region can provide interesting information
about the IBL type BL Lac object 3C~66A, and, during low flux periods of that,
also the FRI type galaxy 3C~66B. 

\acknowledgments
%\input{acknowledgments.tex}
%%%%%%%%%%
%\section{Acknowledgments}
%%%%%%%%%%
We thank the Instituto de Astrof\'{\i}sica de
Canarias for the excellent working conditions at the
Observatorio del Roque de los Muchachos in La Palma.
The support of the German BMBF and MPG, the Italian INFN, 
the Swiss National Fund SNF, and the Spanish MICINN is 
gratefully acknowledged. This work was also supported by 
the Marie Curie program, by the CPAN CSD2007-00042 and MultiDark
CSD2009-00064 projects of the Spanish Consolider-Ingenio 2010
programme, by grant DO02-353 of the Bulgarian NSF, by grant 127740 of 
the Academy of Finland, by the YIP of the Helmholtz Gemeinschaft, 
by the DFG Cluster of Excellence ``Origin and Structure of the 
Universe'', and by the Polish MNiSzW Grant N N203 390834.
The {\it Fermi} data were obtained from the High Energy Astrophysics 
Science Archive Research Center (HEASARC), provided by NASA's 
Goddard Space Flight Center.


%% in parentheses, after the keyword, but they will not be verified.

%{\it Facilities:} \facility{MAGIC}, \facility{{\it Fermi}}, \facility{Milagro}}.

\end{document}